\documentclass[aps,pre,reprint,superscriptaddress]{revtex4-2}

\usepackage{amsmath,amssymb,amsfonts}
\usepackage{graphicx}
\usepackage{hyperref}
\usepackage{xcolor,bm}

\begin{document}

\title{Universal Construction of Generalized Lyapunov Functions for Nonlinear Dynamical Systems Using Physics-Informed Neural Networks}

\author{Z. C. Tu}
\email{tuzc@bnu.edu.cn}
\affiliation{School of Physics and Astronomy, Beijing Normal University, and Key Laboratory of Multiscale Spin Physics (Beijing Normal University), Ministry of Education, Beijing 100875, China}


\begin{abstract}
A scalar potential landscape provides an intuitive description of the stability, transitions, and global organization of dynamical systems. For non-gradient dynamics, however, constructing global Lyapunov-type functions for nonlinear flows with recurrent structures remains a formidable challenge. Here, we introduce the generalized Lyapunov function (GLF), a scalar potential that is monotonically non-increasing along deterministic trajectories, as a unifying framework for nonequilibrium potentials. Conventional Lyapunov functions, Freidlin--Wentzell quasi-potentials, and potentials arising from Ao-type decompositions naturally emerge as special cases within this framework. Furthermore, we develop a data-free physics-informed neural network (PINN) approach in which the Lyapunov inequality and a weak divergence-scale compatibility condition are directly embedded into the loss function. The framework is demonstrated on diverse systems, including linear dynamics, the Hopf normal form, the van der Pol oscillator, a three-dimensional Hopf-link flow, and symmetric and asymmetric competitive Lotka--Volterra systems. The learned landscapes agree closely with analytical results where available and, more generally, successfully reveal invariant sets as low- or constant-potential structures. In particular, the van der Pol oscillator, Hopf-link flow, and asymmetric Lotka--Volterra system demonstrate that coherent GLFs can be constructed without relying on known closed-form expressions, establishing a versatile route to potential-landscape construction for complex nonlinear non-gradient systems.
\end{abstract}

\maketitle

\section{Introduction}
A central goal in the study of nonlinear dynamics is to reduce a complicated vector field to a scalar landscape that reveals where trajectories are attracted, how robust an attractor is, and how different dynamical regions are globally organized. The concept of a potential landscape provides a visual and mathematical representation of stability, transition pathways, and robustness, and has been widely used in control theory, gene regulatory networks, and ecological evolution~\cite{WangJ2015,QianH2017,Maschke1998,Cheng2000,ZhuAo2004,Feinberg2023}. However, most physical and biological systems of current interest are not gradient systems. They often contain rotational currents, dissipative and conservative components, limit cycles, and multiple invariant sets. In such cases, the scalar potential obtained from the Helmholtz decomposition~\cite{Helmholtz1867,VonWestenholz-book} is generally not a suitable global landscape function~\cite{WangJ2015,QianH2017}.

Alternatively, Lyapunov functions offer a powerful approach to certifying stability without explicitly solving the equations of motion~\cite{Lyapunov1892,Khalil2002}. Yet, for nonlinear non-gradient systems, their construction remains highly problem-dependent. While Zubov's equation provides a rigorous theoretical route to Lyapunov functions~\cite{Zubov64}, and approaches based on the Freidlin--Wentzell large deviation theory~\cite{FreidlinWentzell2012} and the Ao decomposition~\cite{Ao2004,Ao2008,Ao2013,Yuan2014} define valuable alternative nonequilibrium potentials, explicit analytical formulas remain elusive beyond special cases. Substantial computational progress has been made via sum-of-squares programming~\cite{Papachristodoulou2002,Papachristodoulou2012}, fractional polynomials~\cite{Vannelli1985}, radial basis functions~\cite{Giesl2007}, and various neural network architectures~\cite{Richards2018,AbateCSL2021,Gruene2021,LiuCDC2023,FengJ2024,Alfarano2024,LiuAutomatica2025}; however, most of these approaches have been developed primarily for isolated
equilibria and their domains of attraction, rather than for recurrent
invariant structures such as periodic orbits or linked cycles.
Even though physics-informed neural networks (PINNs) have successfully uncovered nontrivial, unstable solutions of nonlinear dynamics that elude conventional numerical schemes~\cite{Wang2023PRL}, there remains a need for a practical computational framework capable of constructing global scalar landscapes for nonlinear systems with complex
recurrent invariant structures.

Conceptually, such a framework requires a global potential that reduces to the conventional scalar potential for gradient flows, unifies classical Lyapunov functions and nonequilibrium quasi-potentials, and remains constructible when analytical expressions are unavailable. Conley's fundamental theorem~\cite{Conley1978,Norton1995} along with its Morse--Smale special case~\cite{Smale1967,PalisDeMelo1982} offers a natural resolution by identifying the ``complete Lyapunov function" as a global scalar object that remains constant on the chain recurrent set and decreases outside it. Nevertheless, these classic topological results are purely existential, leaving a conspicuous methodological gap between theoretical existence and explicit numerical construction.

In this work, we address this gap by introducing a generalized Lyapunov function (GLF) as a broad candidate for a nonequilibrium potential and by developing a universal construction based on PINNs. The core idea is to train a neural scalar field under two physics-based constraints: a Lyapunov inequality that enforces non-increase along the flow, and a weak divergence-scale compatibility condition that selects a physically meaningful representative from the highly nonunique family of admissible GLFs. The resulting framework is data-free, uses automatic differentiation to evaluate the required derivatives, and does not assume that the vector field is close to a gradient field. We show that the method recovers known analytical landscapes for two-dimensional (2D) linear systems, the Hopf normal form, and the three-dimensional (3D) symmetric May--Leonard system, while producing coherent potential landscapes for the van der Pol oscillator, 3D Hopf-link system with two linked limit cycles, and an asymmetric competitive Lotka--Volterra system without relying on known closed-form GLFs. These examples demonstrate that GLFs can identify invariant sets as low- or constant-potential structures and provide a practical landscape description for nonlinear non-gradient dynamics even when an explicit analytical GLF is unavailable.

\section{Generalized Lyapunov Functions}
Consider an autonomous dynamical system
\begin{equation}
\dot{\mathbf{x}}=\mathbf{f}(\mathbf{x}),
\end{equation}
where $\mathbf{x}$ represents a phase point, and $\mathbf{f}$ is a smooth vector field. A classical Lyapunov function is usually required to be positive definite with respect to a chosen stable invariant set. For the purpose of constructing a global landscape, we retain only the monotonicity requirement and define a generalized Lyapunov function (GLF) $\phi(\mathbf{x})$ by
\begin{equation}\label{eq-Lyapcond}
\dot{\phi}=\nabla\phi\cdot\mathbf{f}(\mathbf{x})\le 0.
\end{equation}
We call Eq.~(\ref{eq-Lyapcond}) the Lyapunov inequality. It states that trajectories can move only downhill or along level sets of $\phi$. Conley's fundamental theorem~\cite{Conley1978,Norton1995} and its Morse--Smale special case~\cite{Smale1967,PalisDeMelo1982} provide the conceptual basis for such a global decreasing function, although an explicit representative is rarely known and is difficult to construct for nonlinear non-gradient flows.

The above definition of GLF unifies several familiar scalar objects. Conventional Lyapunov functions, Freidlin--Wentzell quasi-potentials, and potentials obtained from the Ao decomposition all satisfy Eq.~(\ref{eq-Lyapcond}) and are therefore GLFs. The definition is intentionally broader than the conventional one: it does not require positive definiteness, a prescribed lower bound, or a unique normalization. Consequently, GLFs are highly nonunique. If $\phi$ is a GLF, then $g(\phi)$ is also a GLF for any smooth increasing function $g$. Moreover, the physical potential of a pure gradient flow is automatically a GLF, and the Helmholtz scalar potential becomes a GLF for gradient-dominant or orthogonal decompositions. Further details and several sufficient conditions for the existence of GLFs are discussed in Secs.~II--IV of the Supplemental Material~\cite{SupplMat}. These observations motivate using Eq.~(\ref{eq-Lyapcond}) as the minimal monotonicity principle and adding a weak selection rule to obtain a physically useful landscape.

\section{Construction via PINNs}
We represent the GLF by a neural network $\phi_\Theta(\mathbf{x})$, where $\Theta$ denotes all trainable network parameters. The method is data-free in the sense that no trajectory data or prescribed potential values are required; the training samples are points in phase space, and the vector field itself supplies the physical constraints. Automatic differentiation provides $\nabla\phi_\Theta$ and $\nabla^2\phi_\Theta$ directly, avoiding finite-difference errors on grids.

The loss function contains three terms. First, the Lyapunov inequality is imposed through
\begin{equation}
\mathcal{L}_\mathrm{Lyap}=\left\langle\mathrm{ReLU}\left(\nabla\phi_\Theta\cdot\frac{\mathbf{f}}{|\mathbf{f}|}\right)\right\rangle,
\end{equation}
which penalizes only points where the learned potential increases along the flow. The normalized direction $\mathbf{f}/|\mathbf{f}|$ gives comparable weight to regions with different flow speeds, especially regions where $|\mathbf{f}|$ is small. Second, the additive constant of the potential is fixed zero by an anchor term at some point $\mathbf{x}_0$,
\begin{equation}
\mathcal{L}_\mathrm{zero}=[\phi_\Theta(\mathbf{x}_0)]^2.
\end{equation}
Third, to avoid the trivial constant solution and to select a physically meaningful representative, we introduce a weak divergence-scale compatibility term
\begin{equation}
\mathcal{L}_\mathrm{div}=\left\langle \left[\frac{\nabla^2\phi_\Theta+\nabla\cdot\mathbf{f}}{1+|\nabla\cdot\mathbf{f}|}\right]^2\right\rangle .\end{equation}
This term does not assume $\mathbf{f}$ to be a gradient field. Instead, it requires the local compression or expansion encoded by $-\nabla\phi_\Theta$ to be compatible in scale with that of the vector field $\mathbf{f}$. For a pure gradient flow $\mathbf{f}=-\nabla U$, the physical potential $\phi=U+C$, with $C$ being a constant, is an exact zero-residual solution of this compatibility condition. For non-gradient flows, the term acts only as a weak scale-selection criterion among the many admissible GLFs; it does not impose a gradient representation of the full vector field.

The total loss is
\begin{equation}\label{eq-loss-tot}
\mathcal{L}_\mathrm{tot}=w_p\mathcal{L}_\mathrm{Lyap}+w_0\mathcal{L}_\mathrm{zero}+w_d\mathcal{L}_\mathrm{div},
\end{equation}
with $w_d\ll w_p\simeq w_0$, so that monotonicity remains the dominant constraint while the divergence term serves only as a weak selector. This structure distinguishes the present framework from a direct regression of $\mathbf{f}\approx-\nabla\phi$, which would incorrectly suppress the rotational component of a non-gradient flow.

We optimize the above loss function using PINNs~\cite{Raissi2019} implemented in PyTorch~\cite{Paszke2019}. The network architectures and training details are provided in Secs.~V and VI of the Supplemental Material~\cite{SupplMat}.

\section{Results}
We apply the above framework to five representative classes of systems of
increasing complexity: 2D linear foci, the Hopf normal form,
the van der Pol oscillator, 3D Hopf-link flow, and
3D competitive Lotka--Volterra systems.

\subsection{Linear systems}

\begin{figure}[!htp]
  \centering
  \includegraphics[width=8.5cm]{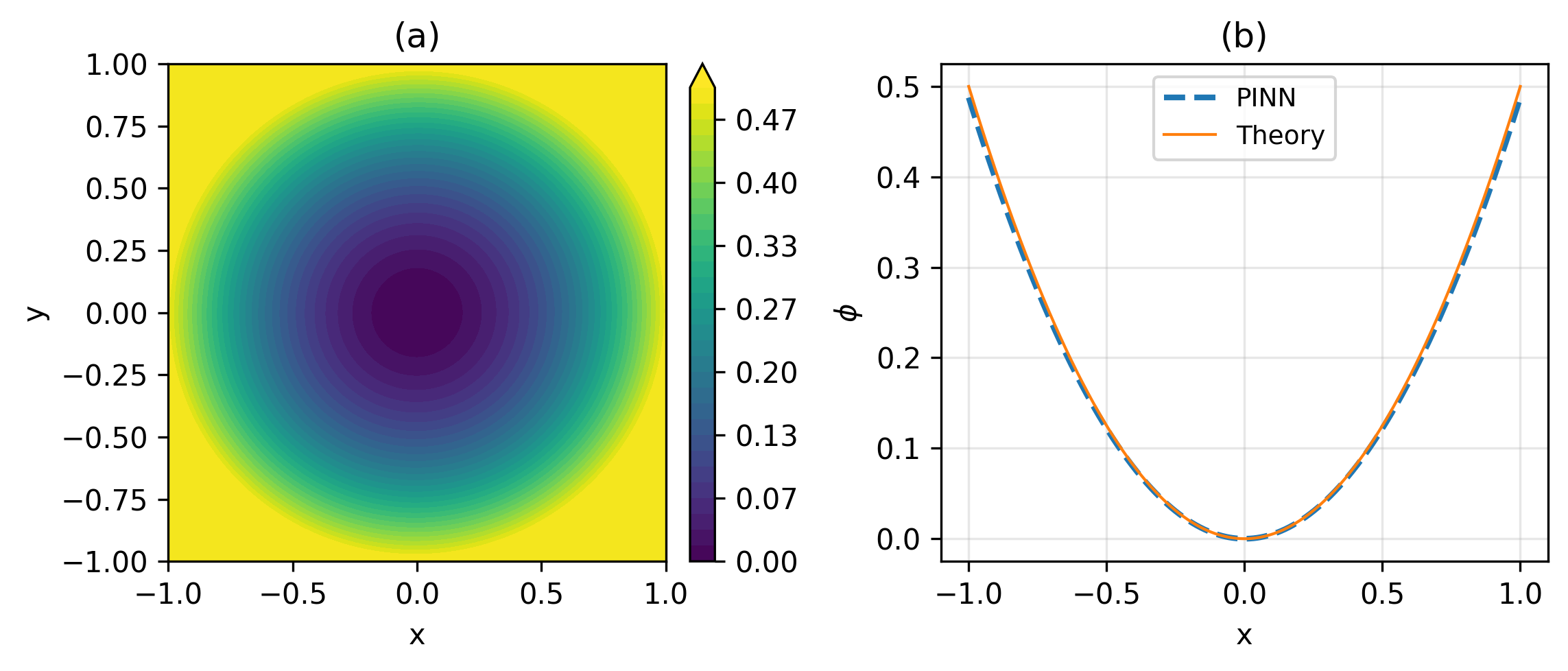}
  \caption{(Color online) GLF for a stable focus: (a) contour plot of PINN-generated GLF; (b) learned and theoretical potentials along the $x$ axis.}\label{fig-linearsy}
\end{figure}

We first validate the framework on 2D linear systems. Figure~\ref{fig-linearsy}a illustrates the contour plot of the PINN-generated GLF for a stable focus governed by $\dot{x}=-x-y$ and $\dot{y}=x-y$, showing excellent agreement with the analytical potential $\phi=(x^2+y^2)/2$. This quantitative match is further confirmed by comparing their profiles along the $x$ axis [Fig.~\ref{fig-linearsy}b]. An analogous agreement is obtained for an unstable focus (see Sec.~VI of the Supplemental Material).

\subsection{Hopf normal form}
A more challenging test involves systems with limit cycles. A solvable benchmark is the 2D Hopf normal form, described by $\dot{x}=x-y-x(x^2+y^2)$ and
$\dot{y}=x+y-y(x^2+y^2)$. A natural exact GLF is
\begin{equation}\label{eq-hopfloop}
\phi_\mathrm{Hopf} = \frac{1}{4}(x^2+y^2-1)^2-\frac{1}{4},
\end{equation}
which has a ``Mexican-hat" shape (see Fig.~3 of the Supplemental Material). It is straightforward to verify $\dot{\phi}_\mathrm{Hopf}\le 0$; related discussions based on the Ao decomposition can be found in Ref.~\cite{Yuan2014}.

\begin{figure}[!htp]
  \centering
  \includegraphics[width=8.5cm]{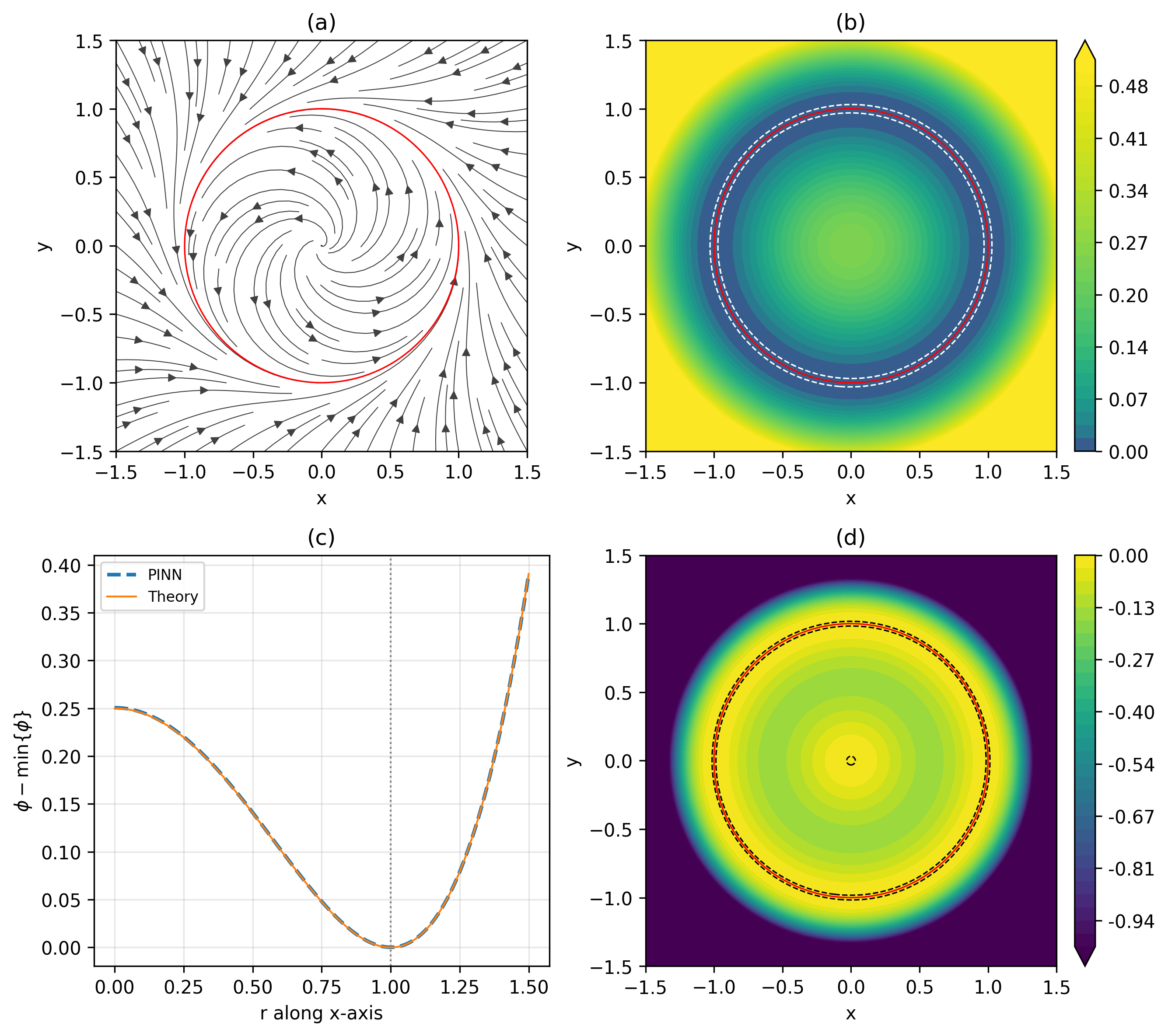}
  \caption{(Color online) Hopf normal form: (a) flow field (arrows) and the unit limit cycle; (b) potential landscape with the unit limit cycle and dashed contours marking $\phi-\min\{\phi\}=0.001$; (c) radial profiles of the PINN-derived potential (dashed line) and the exact potential (solid line); (d) directional derivative $\nabla\phi\cdot\mathbf{f}$, together with the limit cycle (unit circle) and the level set $\nabla\phi\cdot\mathbf{f}=-0.001$ marked by dashed curves.}\label{fig-hopfloop}
\end{figure}

The results obtained from the PINN construction are shown in Fig.~\ref{fig-hopfloop}. The flow field is shown in Fig.~\ref{fig-hopfloop}a, and the landscape $\phi-\min\{\phi\}$ in Fig.~\ref{fig-hopfloop}b reveals the characteristic ``Mexican-hat" geometry. The limit cycle (unit circle) is drawn only as a reference; it was not imposed as a supervised target during training. The dashed contours mark $\phi-\min\{\phi\}=0.001$ and bound the low-potential annulus around the attracting cycle. Figure~\ref{fig-hopfloop}c compares the radial profile of the learned potential with the exact potential derived from (\ref{eq-hopfloop}), showing excellent agreement. The directional derivative $\nabla\phi\cdot\mathbf{f}$ is shown in Fig.~\ref{fig-hopfloop}d, together with the level set $\nabla\phi\cdot\mathbf{f}=-0.001$ marked by dashed curves. We observe that $\nabla\phi\cdot\mathbf{f}$ is nearly zero at the unstable fixed point $(0,0)$ and in the vicinity of the limit cycle, while remaining negative elsewhere. This behavior is consistent with Conley's fundamental theorem: the scalar potential is constant on the chain recurrent set and decreases along trajectories outside it~\cite{Conley1978,Norton1995}.

\begin{figure*}[!htp]
  \centering
  \includegraphics[width=\textwidth]{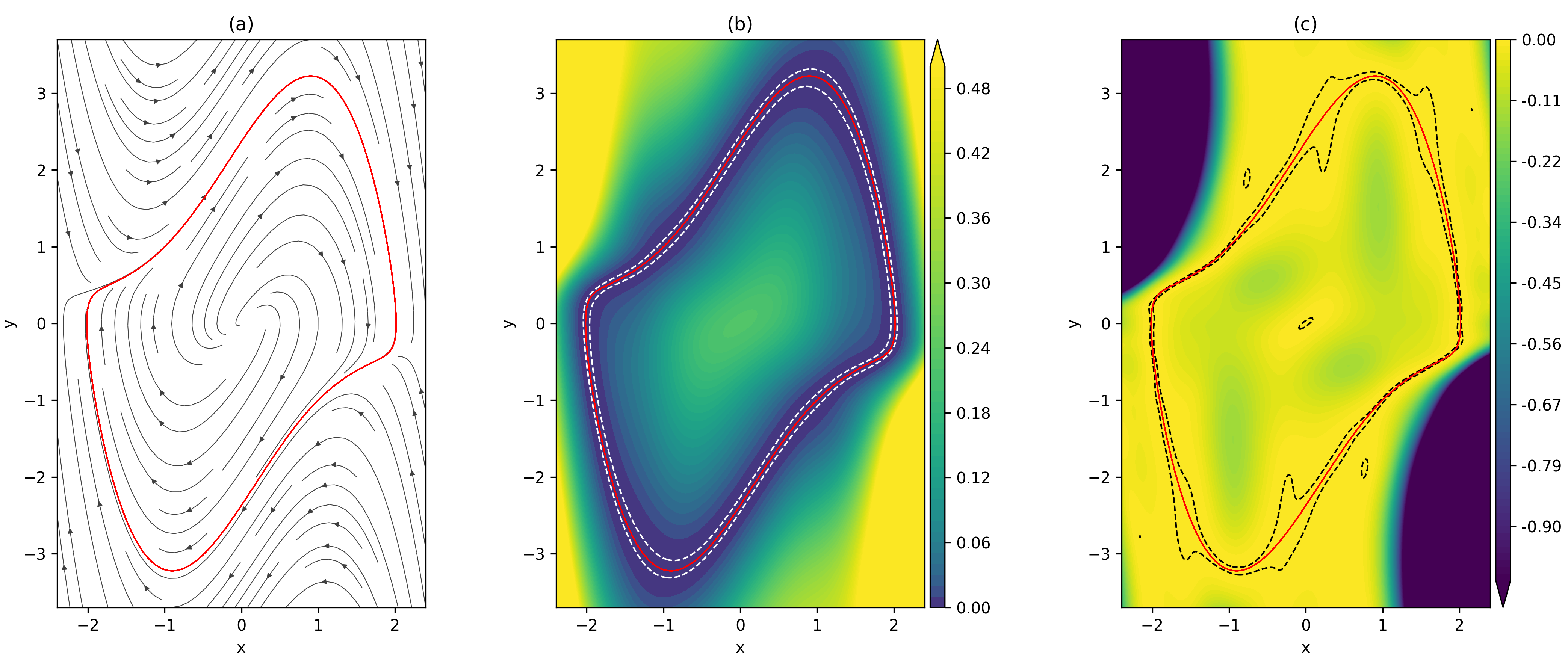}
\caption{(Color online) Van der Pol oscillator: (a) flow field (arrows) and the limit cycle (solid curve); (b) potential landscape with the limit cycle (solid curve) and dashed contours marking the level set $\phi-\min{\phi}=0.001$; (c) directional derivative $\nabla\phi\cdot\mathbf{f}$ with the limit cycle (solid curve) and dashed curves marking the level set $\nabla\phi\cdot\mathbf{f}=-0.0005$.}
\label{fig-vdp}
\end{figure*}

\subsection{van der Pol oscillator}

Here we investigate the van der Pol oscillator, a classical non-conservative system with a stable limit cycle. Compared with the Hopf normal form, this example is more challenging because its limit cycle is strongly noncircular and no simple analytical potential is known for the standard system. Yuan \textit{et al.} have studied a modified van der Pol type oscillator with an analytically tractable potential~\cite{Ao2013}; here we still focus on the standard form
\begin{equation}
    \dot{x}=y,\qquad \dot{y}=-x+\mu(1-x^2)y.
\end{equation}

The flow field and numerically computed limit cycle with $\mu=1.5$ are shown in Fig.~\ref{fig-vdp}a. The learned landscape in Fig.~\ref{fig-vdp}b places the limit cycle (solid curve) inside the low-potential valley, as expected for an attracting periodic orbit. The dashed contours indicate the level set $\phi-\min\{\phi\}=0.001$, providing a local geometric envelope of the learned attractor. The directional derivative $\nabla\phi\cdot\mathbf{f}$ is shown in Fig.~\ref{fig-vdp}c, where the dashed curves indicate the level set $\nabla\phi\cdot\mathbf{f}=-0.0005$. We observe that $\nabla\phi\cdot\mathbf{f}$ is close to zero near the unstable fixed point $(0,0)$ and the stable limit cycle, while remaining negative in the surrounding region. This behavior is consistent with the qualitative properties of a complete Lyapunov function in Conley's fundamental theorem.

\subsection{Hopf-link flow}
\begin{figure}[!htp]
  \centering
  \includegraphics[width=7cm]{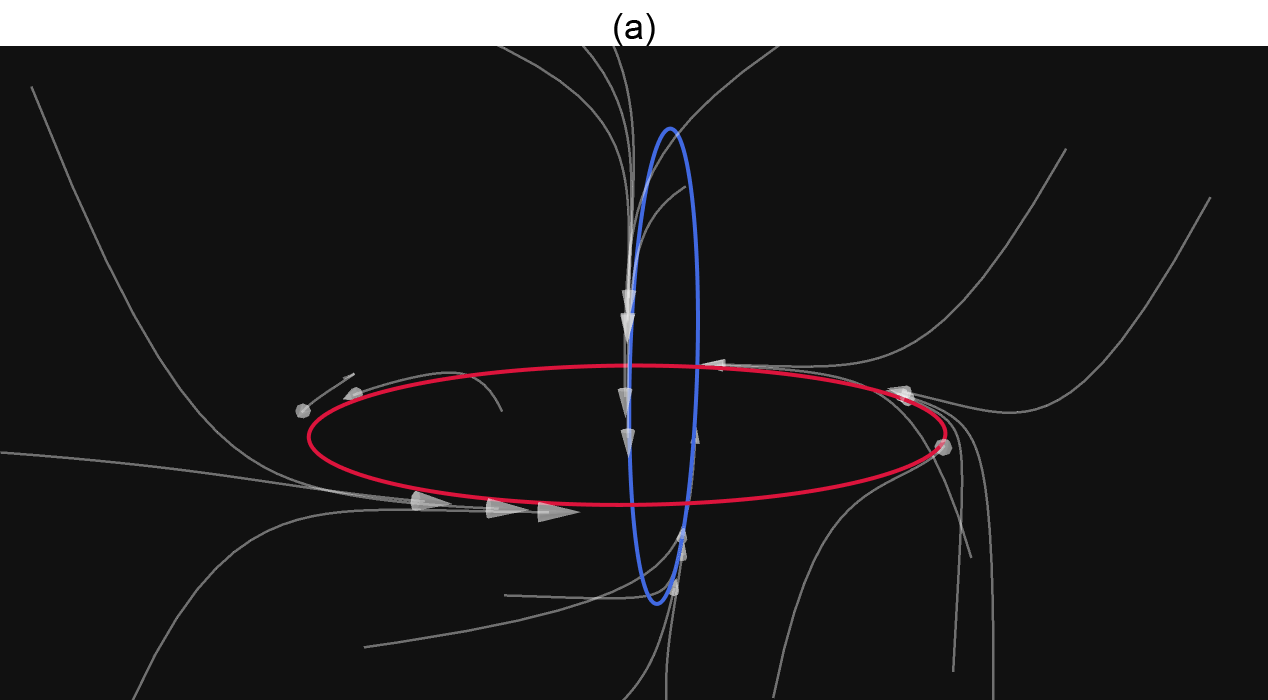}\\
  \includegraphics[width=8.5cm]{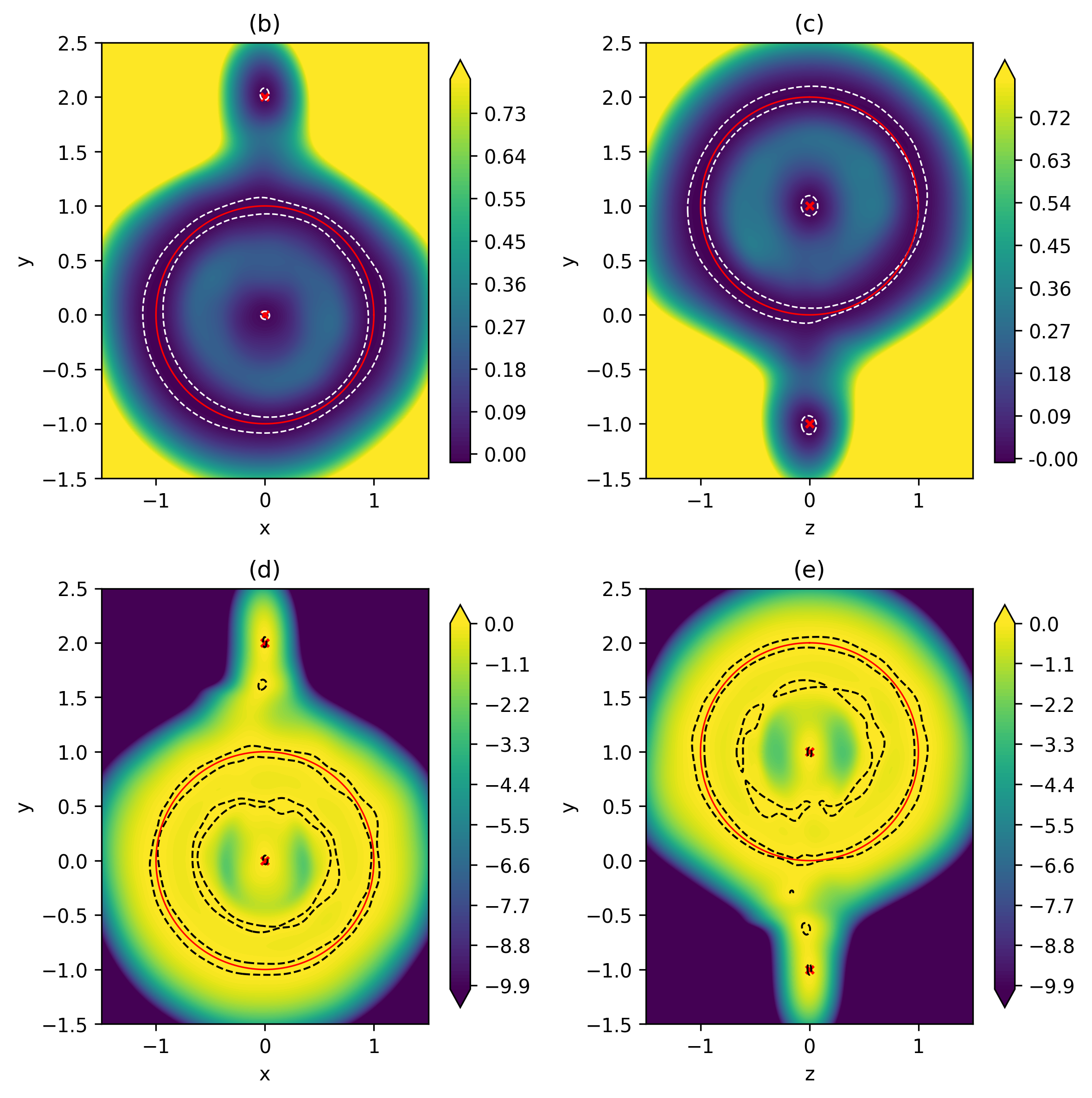}
\caption{(Color online) Hopf-link flow: (a) flow field (arrows) near the two linked limit cycles (solid curves); (b,c) potential landscapes in the $xy$ and $yz$ slices, respectively, with one limit cycle shown as a solid curve, the intersections of the other limit cycle with each slice marked by crosses, and dashed contours indicating the level set $\phi-\min{\phi}=0.02$; (d,e) directional derivatives $\nabla\phi\cdot\mathbf{f}$ in the $xy$ and $yz$ slices, respectively, with the same geometric markers and dashed contours indicating the level set $\nabla\phi\cdot\mathbf{f}=-0.02$.}
\label{fig-hopflink}
\end{figure}

A more complicated test is a 3D system in which topology plays an essential role. We consider a Hopf-link flow described by $\dot{\mathbf{x}}=w\mathbf{F}_1+(1-w)\mathbf{F}_2$, where
\begin{equation}\label{eq-field1}
\mathbf{F}_1=\begin{pmatrix} -y + x(1-x^2-y^2) \\ x + y(1- x^2-y^2) \\ - 10z \end{pmatrix},
\end{equation}
and
\begin{equation}\label{eq-field2}
\mathbf{F}_2=\begin{pmatrix} -10 x \\ - z + (y-1)[1-(y-1)^2-z^2] \\ y-1 + z[1-(y-1)^2-z^2]\end{pmatrix}.
\end{equation}
The distance-dependent mixing factor is chosen as $w={e^{-8d_1}}/({e^{-8d_1}+e^{-8d_2}})$
with $d_1=(\sqrt{x^2+y^2}-1)^2+z^2$ and $d_2=x^2+[\sqrt{(y-1)^2+z^2}-1]^2$. The system possesses two stable limit cycles, $x^2+y^2=1$ in the $xy$ plane and $(y-1)^2+z^2=1$ in the $yz$ plane, which form a linked pair. The local flow near the two limit cycles is shown in Fig.~\ref{fig-hopflink}a. The learned potential landscapes in the $xy$ and $yz$ slices are shown in Figs.~\ref{fig-hopflink}b and c, respectively. The linked cycles appear as low-potential structures, while the contours $\phi-\min{\phi}=0.02$ delineate cross-sections of tube-like neighborhoods surrounding them. The corresponding directional derivatives $\nabla\phi\cdot\mathbf{f}$ are shown in Fig.~\ref{fig-hopflink}d and e. The dashed contours surrounding the two stable limit cycles, together with the inner dashed structures associated with neighborhoods of unstable equilibria, all represent the level set $\nabla\phi\cdot\mathbf{f}=-0.02$. Away from these structures, $\nabla\phi\cdot\mathbf{f}$ takes more negative values. This behavior is again consistent with the qualitative properties of the complete Lyapunov function in Conley's fundamental theorem. This example shows that the GLF construction is not restricted to 2D systems but can also represent global landscape structures associated with nontrivial topology.

\subsection{Lotka--Volterra systems}

We next consider the three-dimensional competitive Lotka--Volterra system~\cite{Lotka1920,Volterra1926,Chi-Wu-Hsu1998},
\begin{equation}\label{eq-Lotka-Volterra}
\left\{
\begin{array}{l}
\dot{x}=x(1-x-\alpha_1 y-\beta_1 z),\\
\dot{y}=y(1-\beta_2 x-y-\alpha_2 z),\\
\dot{z}=z(1-\alpha_3 x-\beta_3 y-z),
\end{array}
\right.
\end{equation}
where $x$, $y$, and $z$ are nonnegative variables, and $\alpha_i$ and $\beta_i$ ($i=1,2,3$) are six interaction parameters. Setting
$\alpha_1=\alpha_2=\alpha_3=\alpha$ and
$\beta_1=\beta_2=\beta_3=\beta$
reduces Eq.~(\ref{eq-Lotka-Volterra}) to the classical May--Leonard system~\cite{May-Leonard1975}.

For the special case $\alpha+\beta=2$, introducing
\begin{equation}
S=x+y+z>0,
\end{equation}
one readily obtains
\begin{equation}
\dot{S}=S(1-S).
\end{equation}
Hence, the simplex $S=1$ is invariant and attracts all trajectories with $S(0)>0$. On this invariant simplex, the May--Leonard system possesses a one-parameter family of periodic orbits surrounding the coexistence equilibrium~\cite{Hirsch1988}. This radial relaxation suggests the dissipative contribution
\begin{equation}\label{eq-dissipf-R}
R=a(S^2-1)^2,
\end{equation}
where $a>0$.

In addition, for $\alpha+\beta=2$,
\begin{equation}\label{eq-hamilton}
H=\frac{xyz}{S^3}
\end{equation}
is a global first integral in the interior of the positive octant, satisfying
\begin{equation}
\dot{H}=\mathbf{f}\cdot\nabla H=0.
\end{equation}
The dissipative relaxation toward $S=1$ and the conservative motion within $S=1$ can therefore be combined into the GLF
\begin{equation}\label{eq-GLF-May-Leonard}
\phi_{\mathrm{ML}}
=
R+c\left(H-\frac{1}{27}\right),
\end{equation}
where $c$ is a constant. With this normalization,
$\phi_{\mathrm{ML}}=0$ at the coexistence point
$(1/3,1/3,1/3)$. Since $H$ is conserved, the time derivative of GLF is determined entirely by $R$:
\begin{equation}\label{eq-dotMay-Leonard}
\dot{\phi}_{\mathrm{ML}}
=
\dot{R}
=
-4aS^2(S-1)^2(S+1)
\le 0.
\end{equation}
Thus, $\dot{\phi}_{\mathrm{ML}}=0$ on the invariant simplex $S=1$ and is strictly negative for $S>0$, $S\neq1$. The detailed derivation is given in Sec.~VI.E of Supplemental Material. We emphasize that the GLF is nonunique. In particular, Eq.~(\ref{eq-GLF-May-Leonard}) differs from the function
$\phi_{\mathrm{TYM}}=3S-\ln(xyz)$ introduced by Tang, Yuan, and Ma~\cite{TangYuanMaPRE2013}.

\begin{figure}[!htp]
  \centering
  \includegraphics[width=8.5cm]{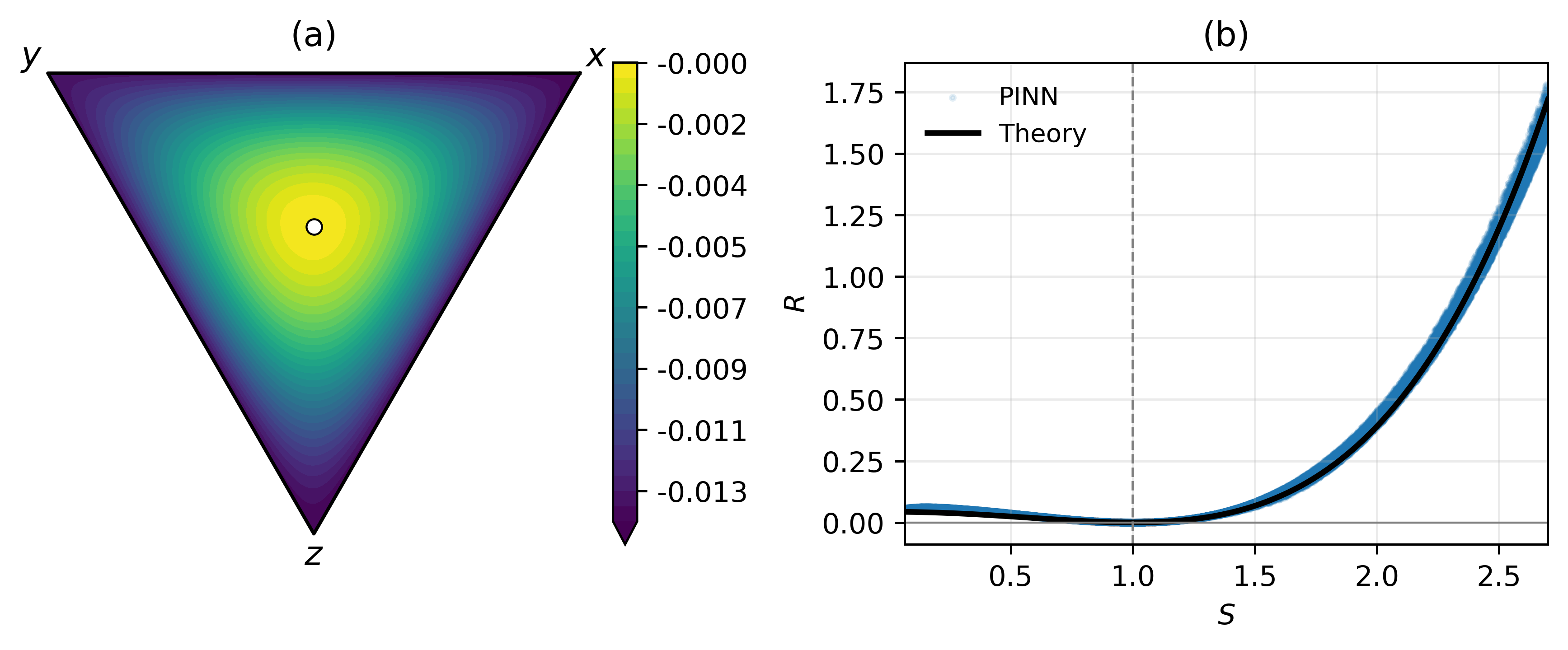}\\
  \includegraphics[width=8.5cm]{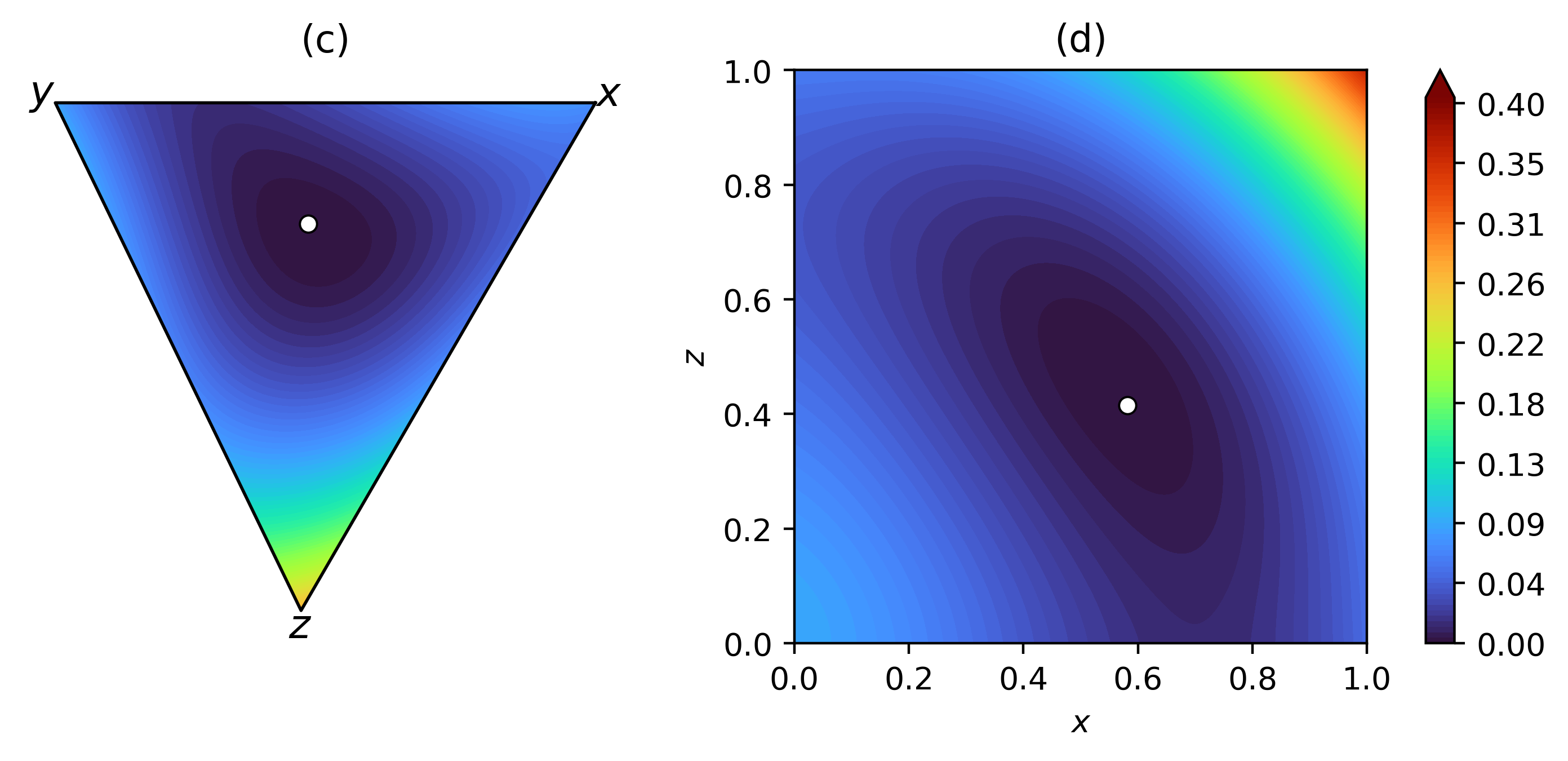}
\caption{(Color online) Lotka--Volterra systems.
(a) Contour plot of the PINN-learned GLF on the invariant simplex $S=1$ of the May--Leonard system.
(b) PINN-learned and theoretical $R$--$S$ relations for the May--Leonard system.
(c) Contour plot of the learned GLF for the asymmetric Lotka--Volterra system on the plane
$0.572x+0.626y+0.530z=0.924$.
(d) Contour plot of the learned GLF on the plane $y=0.593$.}
\label{fig-Lotka-Volterra}
\end{figure}

Remarkably, the PINN-learned potential is accurately described by the analytical form in Eq.~(\ref{eq-GLF-May-Leonard}). For $\alpha=1.20$ and $\beta=0.80$, fitting the numerical result gives
$a=0.0434$ and $c=0.356$, with a coefficient of determination equal to $0.997$. Figure~\ref{fig-Lotka-Volterra}a shows the learned potential on the invariant simplex $S=1$, where the closed contours coincide with level sets of the conserved
quantity $H$ and therefore label the periodic orbits on the invariant
simplex. To isolate the dissipative contribution, we define
\begin{equation}
R_{\mathrm{PINN}}
=
\phi-\phi_c-
c\left(H-\frac{1}{27}\right),
\end{equation}
where $\phi_c$ is the value of GLF at the coexistence point
$(1/3,1/3,1/3)$.
As shown in Fig.~\ref{fig-Lotka-Volterra}b, the PINN result collapses onto the theoretical relation $R=a(S^2-1)^2$ over a broad range of $S$. This agreement shows that the learned GLF simultaneously resolves the attraction toward the invariant simplex and the neutral dynamics within it. In this sense, the present example goes beyond fixed points and isolated limit cycles and demonstrates that the construction can identify a two-dimensional invariant set carrying a continuous family of periodic orbits.

We further break the cyclic symmetry by choosing
$(\alpha_1,\alpha_2,\alpha_3) =(0.60,0.70,0.75)$ and $(\beta_1,\beta_2,\beta_3)=(0.15,0.20,0.25)$.
For this asymmetric Lotka--Volterra system, the interior equilibrium is located at
$\mathbf{x}_c\simeq (0.582,\,0.593,\,0.415)$.
Unlike the symmetric May--Leonard case, the simple decomposition into a scalar relaxation variable $S$ and an independent first integral $H$ is no longer available, and no corresponding closed-form GLF is known for this parameter set. Figures~\ref{fig-Lotka-Volterra}c and \ref{fig-Lotka-Volterra}d show two sections of the PINN-learned potential. The first plane, $0.572x+0.626y+0.530z=0.924$, passes through the coexistence equilibrium and has a normal vector
approximately equal to the left eigenvector associated with the real
eigenvalue $\lambda_1=-1$. It therefore isolates, to linear order, the
two-dimensional spiral subspace associated with the complex-conjugate
eigenvalue pair. The second section is taken at $y=0.593$. The coexistence equilibrium $\mathbf{x}_c$ is indicated by the white circles. In both sections, the learned potential forms a well-defined basin centered at the equilibrium, while the distorted, non-axisymmetric contours reflect the broken cyclic symmetry. Linear stability analysis gives one real negative eigenvalue and a complex-conjugate pair with negative real part, confirming that the coexistence equilibrium is a stable focus (see Sec.~VI.E in the Supplemental Material). The agreement between this dynamical classification and the geometry of the learned potential demonstrates that the PINN construction remains effective even when the symmetry and analytical integrability of the May--Leonard model are removed.

\section{Conclusion}
We have introduced GLFs as a flexible notion of potential for nonlinear non-gradient dynamics and developed a data-free PINN framework for their construction. The significance of the approach is twofold. Conceptually, it places Lyapunov stability, nonequilibrium quasi-potentials, Ao-type potentials, and the global ordering implied by Conley's fundamental theorem into a unified scalar-landscape framework. Computationally, it recasts the search for such a scalar function as the minimization of a physics-informed loss that enforces monotonicity along the flow while employing a weak divergence-scale compatibility condition to select a dynamically meaningful representative from the nonunique family of admissible GLFs.

The numerical examples show that the learned GLFs recover known analytical
potentials for benchmark systems while also revealing invariant structures
without relying on closed-form expressions. For the van der Pol oscillator
and the Hopf-link flow, the learned landscapes identify stable periodic
structures as low-potential regions. In the symmetric May--Leonard system,
the PINN resolves both the dissipative relaxation toward the invariant
simplex and the neutral periodic dynamics within it, in quantitative
agreement with the analytical GLF derived here. After the cyclic symmetry
is broken, the asymmetric Lotka--Volterra system still exhibits a coherent
asymmetric potential basin consistent with the independently
determined stable-focus dynamics. These results suggest that GLFs provide
a practical landscape framework for investigating the global organization
of nonlinear non-gradient systems with increasingly complex invariant
structures.

It is also useful to contrast the present approach with recent symbolic-learning methods for Lyapunov-function construction. In particular,
Alfarano \textit{et al.}~\cite{Alfarano2024} employ a symbolic transformer to generate explicit analytical Lyapunov functions from the equations of motion. Such an approach has the important advantage that, when a compact symbolic expression can be identified, the resulting function is directly interpretable and can potentially be subjected to symbolic or formal verification. The objective of the present framework is complementary. Rather than assuming that the GLF admits a compact closed-form representation, we construct it directly as a continuous neural scalar field subject to dynamical constraints. This is particularly useful when a compact closed-form representation of the GLF is unavailable, as illustrated by the van der Pol, Hopf-link, and asymmetric Lotka--Volterra examples, and for systems involving periodic orbits, multiple recurrent sets, or higher-dimensional invariant structures. The price for this flexibility is that the PINN-derived GLF is usually numerical and requires system-specific training and subsequent validation. Symbolic-transformer and PINN-based approaches therefore address different aspects of the Lyapunov-function construction problem and may be viewed as complementary rather than competing strategies.

The present framework suggests several avenues for further theoretical analysis and computational extension. First, the mathematical mechanism by which the PINN loss selects a dynamically meaningful representative from the highly nonunique family of GLFs deserves further investigation. In particular, it is important to establish sufficient conditions under which the Lyapunov inequality together with the weak divergence-scale compatibility condition yields landscape structures consistent with the qualitative properties of complete Lyapunov functions in Conley's fundamental theorem. Second, the mathematical relation between the present construction and the Ao decomposition remains subtle. Although both frameworks lead to scalar functions that decrease along deterministic trajectories, the weak divergence-scale compatibility condition introduced here is not fully equivalent to the tensorial structure assumed in the Ao decomposition. Clarifying whether, and under what additional conditions, the two formulations can be related or partially mapped onto each other is a challenging problem that we leave for future work. Finally, extending the present method to chaotic attractors and high-dimensional systems is an important computational direction. To mitigate the curse of dimensionality, one may cover a large phase-space domain by an atlas of overlapping local computational charts. A local GLF can then be constructed on each chart, with consistency conditions imposed on the overlaps to assemble the local solutions into a global potential landscape.

\section*{Acknowledgments}
The author is grateful to Ping Ao and Haiwen Liu for drawing his attention to the problem of constructing potential landscapes. The author thanks Ping Ao, Gang Hu, Hong Zhao, Weimou Zheng, and Haijun Zhou for useful discussions. This work was supported by the National Natural Science Foundation of China (Grant Nos.~12475032 and 11975050). Some of the computational resources were kindly provided by Jing Wu. The author used ChatGPT for language polishing, coding, and debugging assistance. All scientific content, numerical methods, and interpretations of the results were checked and finalized by the author.


\begin{thebibliography}{99}
\bibitem{Maschke1998}B. M. Maschke, R. Ortega, and A. J. van der Schaft, Energy-based Lyapunov functions for forced Hamiltonian systems with dissipation, Proceedings of the 37th IEEE Conference on Decision and Control (1998).
\bibitem{Cheng2000}D. Cheng, S. Spurgeon, and J. Xiang, On the development of generalized Hamiltonian realizations, Proceedings of the 39th IEEE Conference on Decision and Control (2000).
\bibitem{ZhuAo2004}X.-M. Zhu, L. Yin, L. Hood, and P. Ao, Robustness, stability and efficiency of phage $\lambda$ genetic switch: dynamical structure analysis, J. Bioinf. Comput. Biol. \textbf{2}, 785 (2004).
\bibitem{Feinberg2023}A. P. Feinberg and A. Levchenko, Epigenetics as a mediator of plasticity in cancer, Science \textbf{379}, eaaw3835 (2023). 

\bibitem{WangJ2015}J. Wang, Landscape and flux theory of non-equilibrium dynamical systems with application to biology, Adv. Phys. \textbf{64}, 1 (2015).
\bibitem{QianH2017}P. Ao, C. Kwon, and H. Qian, On the existence of potential landscape in the evolution of complex systems, Complexity \textbf{12}, 19 (2007).

\bibitem{Helmholtz1867}H. Helmholtz, LXIII. On Integrals of the hydrodynamical equations, which express vortex-motion. The London, Edinburgh, and Dublin Philosophical Magazine and Journal of Science \textbf{33}, 485 (1867).
\bibitem{VonWestenholz-book}C. Von Westenholz, \textit{Differential Forms in Mathematical Physics} (Elsevier Science, 2009).

\bibitem{Lyapunov1892}A. M. Lyapunov, The general problem of the stability of motion (Translated from French into English by A. T. Fuller). Int. J. Control \textbf{55}, 531 (1992).
\bibitem{Khalil2002}H. K. Khalil, \textit{Nonlinear Systems}, 3rd ed. (Prentice Hall 2002).


\bibitem{Zubov64}V. I. Zubov. \textit{Methods of A. M. Lyapunov and Their Application} (Noordhoff, Groningen, 1964).
\bibitem{FreidlinWentzell2012}M. I. Freidlin and A. D. Wentzell, \textit{Random Perturbations of Dynamical Systems}, 3rd ed. (Springer, Heidelberg, 2012).
\bibitem{Ao2004}P. Ao, Potential in stochastic differential equations: novel construction, J. Phys. A: Math. Gen. \textbf{37}, L25 (2004).
\bibitem{Ao2008}P. Ao, Emerging of Stochastic Dynamical equalities and steady state thermodynamics from darwinian dynamics, Commun. Theor. Phys. \textbf{49}, 1073 (2008).
\bibitem{Yuan2014}R.-S. Yuan, Y.-A. Ma, B. Yuan, and P. Ao, Lyapunov function as potential function: A dynamical equivalence, Chinese Phys. B \textbf{23}, 010505 (2014).
\bibitem{Ao2013}R. Yuan, X. Wang, Y. Ma, B. Yuan, and P. Ao, Exploring a noisy van der Pol type oscillator with a stochastic approach, Phys. Rev. E \textbf{87}, 062109 (2013).  

\bibitem{Papachristodoulou2002}A. Papachristodoulou and S. Prajna, On the construction of Lyapunov functions using the sum of squares decomposition, Proc. IEEE Conf. Decision Control \textbf{3}, 3482 (2002). 
\bibitem{Papachristodoulou2012}M. M. Peet and A. Papachristodoulou, A converse sum of squares Lyapunov result with a degree bound, IEEE Trans. Automatic Control \textbf{57}, 2281 (2012). 

\bibitem{Vannelli1985}A. Vannelli and M. Vidyasagar, Maximal lyapunov functions and domains of attraction for autonomous nonlinear systems, Automatica \textbf{21}, 69 (1985). 

\bibitem{Giesl2007}P. Giesl, \textit{Construction of Global Lyapunov Functions Using Radial Basis Functions} (Springer, Heidelberg, 2007). 


\bibitem{Richards2018}S. M. Richards, F. Berkenkamp, and A. Krause, The Lyapunov neural network: Adaptive stability certification for safe learning of dynamical systems, Proc. Machine Learning Research \textbf{87}, 466 (2018).
\bibitem{AbateCSL2021}A. Abate, D. Ahmed, M. Giacobbe, and A. Peruffo, Formal synthesis of Lyapunov neural networks, IEEE Control Systems Letters \textbf{5}, 773 (2021).
\bibitem{Gruene2021}L. Gr\"{u}ne, Computing Lyapunov functions using deep neural networks, J. Comput. Dynamics \textbf{8}, 131 (2021).
\bibitem{LiuCDC2023}J. Liu, Y. Meng, M. Fitzsimmons, and R. Zhou, Towards learning and verifying maximal neural Lyapunov functions, 62nd IEEE Conference on Decision and Control (Marina Bay Sands, Singapore, 2023).
\bibitem{FengJ2024}J. Feng, H. Zou, and Y. Shi, Combining neural networks and symbolic regression for analytical Lyapunov function discovery, arXiv.2406.15675 (2024). 
\bibitem{Alfarano2024}A. Alfarano, F. Charton, and A. Hayat, Global Lyapunov functions: a long-standing open problem in mathematics, with symbolic transformers, arXiv:2410.08304 (2024). 
\bibitem{LiuAutomatica2025}J. Liu, Y. Meng, M. Fitzsimmons, and R. Zhou, Physics-informed neural network Lyapunov functions: PDE characterization, learning, and verification, Automatica \textbf{175}, 112193 (2025). 

\bibitem{Wang2023PRL}Y. Wang, C.-Y. Lai, J. G\'omez-Serrano, and T. Buckmaster, Asymptotic self-similar blow-up profile for three-dimensional axisymmetric Euler equations using neural networks, Phys. Rev. Lett. \textbf{130}, 244002 (2023).

\bibitem{Conley1978}C. Conley, \textit{Isolated Invariant Sets and the Morse Index} (American Mathematical Society, 1978).
\bibitem{Norton1995}D.~E. Norton, The fundamental theorem of dynamical systems, Commentationes Mathematicae Universitatis Carolinae \textbf{36}, 585, (1995).

\bibitem{Smale1967}S. Smale, Differentiable dynamical systems, Bulletin Amer. Math. Soc. \textbf{73}, 747, (1967).
\bibitem{PalisDeMelo1982}J. Palis and W. de Melo, \textit{Geometric Theory of Dynamical Systems: An Introduction} (Springer, 1982).

\bibitem{SupplMat}See the Supplemental Material for details on properties of GLFs, network architectures and training hyperparameters. The Supplemental Material may be requested by sending email to tuzc@bnu.edu.cn.

\bibitem{Raissi2019}M. Raissi, P. Perdikaris, and G.~E. Karniadakis, Physics-informed neural networks: A deep learning framework for solving forward and inverse problems involving nonlinear partial differential equations, J. Comput. Phys. \textbf{378}, 686 (2019).


\bibitem{Paszke2019}A. Paszke, S. Gross, F. Massa, A. Lerer, J. Bradbury, G. Chanan, T. Killeen, Z. Lin, N. Gimelshein, L. Antiga, A. Desmaison, A. Kopf, E. Yang, Z. DeVito, M. Raison, A. Tejani, S. Chilamkurthy, B. Steiner, L. Fang, J. Bai, S. Chintala, PyTorch: An imperative style, high-performance deep learning library, Proceedings of the 33rd International Conference on Neural Information Processing Systems, 8026 (2019).

\bibitem{Lotka1920}A. Lotka, Analytical note on certain rhythmic relations in organic systems, Proc. Natl. Acad. Sci. USA \textbf{6}, 410 (1920). 

\bibitem{Volterra1926}V. Volterra, Fluctuations in the abundance of a species considered mathematically, Nature \textbf{118}, 558 (1926).

\bibitem{Chi-Wu-Hsu1998}C.-W. Chi, L.-I. Wu, and S.-B. Hsu, On the asymmetric May-Leonard model of three competing species, SIAM J. Appl. Math. \textbf{58}, 211 (1998). 

\bibitem{May-Leonard1975}R. May and W. Leonard, Nonlinear aspects of competition between three species, SIAM J. Appl. Math. \textbf{29}, 243 (1975). 

\bibitem{Hirsch1988}M. Hirsch, Systems of differential equations which are competitive or cooperative: III. Competing species, Nonlinearity \textbf{1}, 51 (1988). 

\bibitem{TangYuanMaPRE2013}Y. Tang, R. Yuan, and Y. Ma, Dynamical behaviors determined by the Lyapunov function in competitive Lotka-Volterra systems, Phys. Rev. E \textbf{87}, 012708 (2013). 

\end{thebibliography}
\end{document}